\documentclass{PoS}

\usepackage{graphicx}
\usepackage{amsmath}
\usepackage{aas_macros}
\usepackage{color}
\usepackage{xcolor}
\usepackage{accents}
\usepackage[procnames]{listings}
\usepackage{listings}

\RequirePackage[left]{lineno}

\definecolor{keywords}{HTML}{007021}
\definecolor{comments}{RGB}{0,0,113}
\definecolor{strings}{HTML}{4070A1}
\definecolor{package}{HTML}{0C85B5}
\definecolor{lightgrey}{rgb}{0.98,0.98,0.98}
\definecolor{numbers}{HTML}{A30000}

\newcommand{\url}[1]{\href{#1}{#1}}

\title{Fermipy: An open-source Python package for analysis of \Fermi-LAT Data}

\ShortTitle{Fermipy: An open-source Python package for analysis of Fermi-LAT Data}

\author{\speaker{Matthew Wood}\\
  Kavli Institute for Particle Astrophysics and Cosmology, SLAC National Accelerator Laboratory\\
  E-mail: \email{mdwood@slac.stanford.edu}}
\author{Regina Caputo\\
  NASA, GSFC\\
  E-mail: \email{regina.caputo@nasa.gov}}
\author{Eric Charles\\
  Kavli Institute for Particle Astrophysics and Cosmology, SLAC National Accelerator Laboratory\\
  E-mail: \email{echarles@slac.stanford.edu}}
\author{Mattia Di Mauro\\
  Kavli Institute for Particle Astrophysics and Cosmology, SLAC National Accelerator Laboratory\\
  E-mail: \email{mdimauro@slac.stanford.edu}}
\author{Jeffrey Magill\\
  Department of Physics and Department of Astronomy, University of Maryland, College Park\\
  E-mail: \email{jmagill@umd.edu}}
\author{Jeremy Perkins\\
  NASA, GSFC\\
  E-mail: \email{jeremy.s.perkins@nasa.gov}}

\author{on behalf of the \Fermi-LAT Collaboration}

\abstract{  
  {\fermipy} is an open-source python framework that facilitates analysis
  of data collected by the {\Fermi} Large Area Telescope (LAT). {\fermipy}
  is built on the {\Fermi} {\stools}, the publicly available software
  suite provided by NASA for the LAT mission. {\fermipy} provides a
  high-level interface for analyzing LAT data in a simple and
  reproducible way. The current feature set includes methods for
  extracting spectral energy distributions and lightcurves, generating
  test statistic maps, finding new source candidates, and fitting
  source position and extension. {\fermipy} leverages functionality from
  other scientific python packages including NumPy, SciPy, Matplotlib,
  and Astropy and is organized as a community-developed package
  following an open-source development model. We review the current
  functionality of {\fermipy} and plans for future development.}

\FullConference{35th International Cosmic Ray Conference -ICRC217-\\
		10-20 July, 2017\\
		Bexco, Busan, Korea}

\newcommand{\stools}{\emph{ScienceTools}}
\newcommand{\Fermi}{{\textit{Fermi}}}

\newcommand{\fermipy}{\emph{Fermipy}}
\newcommand{\pylike}{\texttt{pyLikelihood}}
\newcommand{\version}{0.14.1}
\newcommand{\findsources}{\href{http://fermipy.readthedocs.org/en/\version/advanced/detection.html}{\lstinline$find_sources$}}
\newcommand{\tsmap}{\href{http://fermipy.readthedocs.org/en/\version/advanced/tsmap.html}{\lstinline$tsmap$}}
\newcommand{\extension}{\href{http://fermipy.readthedocs.org/en/\version/advanced/extension.html}{\lstinline$extension$}}
\newcommand{\sed}{\href{http://fermipy.readthedocs.org/en/\version/advanced/sed.html}{\lstinline$sed$}}
\newcommand{\localize}{\href{http://fermipy.readthedocs.org/en/\version/advanced/localization.html}{\lstinline$localize$}}

\begin{document}

\section{Introduction}

The {\Fermi} Large Area Telescope (LAT) is a pair-conversion telescope
that is sensitive to gamma rays from below 20 MeV to more than 300
GeV.  Since the launch of the {\Fermi} observatory in June 2008, the
LAT has operated primarily in all-sky survey mode and has now
collected more than eight years of data.  LAT data classified as
photon-like (satisfying the criteria of the photon event classes) are
immediately released to the public
%LAT data are distributed to the public 
through the LAT data
archive\footnote{\url{https://fermi.gsfc.nasa.gov/cgi-bin/ssc/LAT/LATDataQuery.cgi}}
at the Fermi Science Support Center (FSSC).  The FSSC also supports a
suite of public software tools, the Fermi {\stools}
\footnote{\url{https://fermi.gsfc.nasa.gov/ssc/data/analysis/software/}},
for the reduction and analysis of LAT data.  The {\stools} is
primarily written in C++ but includes a python interface ({\pylike}) to
facilitate scripting analysis in python.

{\fermipy} is a python software package that provides a high-level
interface for LAT data analysis.  {\fermipy} is based on the {\stools}
and uses {\pylike} to interface with the underlying C++ library.
{\fermipy} relies on a number of other open-source python libraries
including NumPy~\cite{oliphant_guide_2006},
Scipy~\cite{scipy}, and
Astropy~\cite{2013A&A...558A..33A}.
Matplotlib~\cite{Hunter:2007:MGE:1251563.1251845} is an optional
dependency that is used to generate analysis visualizations.

{\fermipy} is organized around an open-source development model and is
available to all members of the LAT scientific community.  The
{\fermipy} source code is hosted on
GitHub\footnote{\url{https://github.com/Fermipy/fermipy}} and is
licensed under a 3-clause BSD-style license.  Bug reports and
proposals for new functionality should be made through the GitHub
issue tracker.  Code contributions are accepted via pull requests.

This proceeding provides a short review of some of the main features
of {\fermipy}.  A more complete and detailed description of the
package as well as tutorials are available in the online
documentation.\footnote{\url{http://fermipy.readthedocs.org/}}
%Section~\ref{sec:analysis}

\section{Installation}

Releases of {\fermipy} are made available through both the
\href{https://pip.pypa.io/en/latest/}{pip} and
\href{https://conda.io/docs/index.html}{conda} package management
tools.  Instructions for pip- and conda-based installation are
provided in the online documentation.
\footnote{\url{http://fermipy.readthedocs.org/en/\version/install.html}}
Both installation methods require an existing installation of
{\stools} provided as precompiled binaries from the FSSC.  Pre-built
\href{https://www.docker.com}{Docker} images containing a current
release of {\fermipy}, the Fermi {\stools}, and Anaconda python are
also available on the Fermipy DockerHub
repository.\footnote{\url{https://hub.docker.com/r/fermipy/fermipy}}

\section{Analysis Workflow}\label{sec:analysis}

{\fermipy} is designed around a global analysis state object
(\href{http://fermipy.readthedocs.io/en/\version/fermipy.html\#fermipy.gtanalysis.GTAnalysis}{GTAnalysis})
that manages the data and model preparation and provides a set of
high-level analysis methods.  A user executes an analysis by composing
a python script that creates a
\href{http://fermipy.readthedocs.io/en/\version/fermipy.html\#fermipy.gtanalysis.GTAnalysis}{GTAnalysis}
instance and calls the methods of this object to perform different
analysis tasks.  A user starts a new analysis by first composing a
configuration file that defines analysis parameters including the data
selection, region-of-interest (ROI) geometry, and model specification.
{\fermipy} uses the YAML format for its configuration files and
organizes parameters in a nested hierarchy that groups related
parameters into sub-dictionaries (e.g. \textit{data},
\textit{binning}, and \textit{selection}).
%groups
%related parameters into a nested hierarchy.  
A sample configuration file is shown
in Figure~\ref{fig:fermipy_config}.  Configurations can easily be
constructed programatically with a YAML serialization and parsing
library such as \href{http://pyyaml.org/wiki/PyYAML}{PyYAML}.

\begin{figure}[t]
\begin{lstlisting}
data:
  evfile : ft1.lst
  scfile : ft2.fits
binning:
  roiwidth   : 10.0
  binsz      : 0.1
  binsperdec : 8
selection :
  emin : 100
  emax : 316227.76
  zmax    : 90
  evclass : 128
  evtype  : 3
  tmin    : 239557414
  tmax    : 428903014
  target : 'mkn421'
gtlike:
  edisp : True
  irfs : 'P8R2_SOURCE_V6'
  edisp_disable : ['isodiff','galdiff']
model:
  src_roiwidth : 15.0
  galdiff  : '$FERMI_DIFFUSE_DIR/gll_iem_v06.fits'
  isodiff  : 'iso_P8R2_SOURCE_V6_v06.txt'
  catalogs : ['3FGL']
\end{lstlisting}
\caption{A YAML configuration for a {\fermipy} analysis of the gamma-ray blazar Mkn~421.\label{fig:fermipy_config}}
\end{figure}

%The user composes an analysis script that creates a
%\href{http://fermipy.readthedocs.io/en/\version/fermipy.html\#fermipy.gtanalysis.GTAnalysis}{GTAnalysis}
%instance and calls the methods of this object.  

Figure~\ref{fig:fermipy_script} shows a sample analysis script that
could be used in conjunction with the configuration file shown in
Figure~\ref{fig:fermipy_config}.  The configuration file sets the
parameters used to initialize the analysis object and its path is
passed to the analysis object constructor.  The \lstinline$setup$
method is called to run the data and model preparation by executing
the appropriate gt-tools: \emph{gtselect} (data selection),
\emph{gtmktime} (data filtering), \emph{gtbin} (data binning),
\emph{gtltcube} (livetime calculation), \emph{gtexpcube2} (exposure
calculation), and \emph{gtsrcmaps} (calculation of spatial templates
for individual model components).

Once the analysis object is initialized, a typical analysis sequence
involves optimizing the parameters of the background model, generating
a Test Statistic (TS) or residual map of the region to assess the
quality of the model, and extracting the characteristics of a source
of interest (flux, TS, spectral fit parameters, SED, etc.).  In the
example script shown in Figure~\ref{fig:fermipy_script}, the
\lstinline$optimize$ method is used to find best-fit values for the spectral parameters of all components of the model.
%model for the
%region by fitting the parameters of all components in the ROI.  
The {\localize} and {\sed} methods are then used to determine the
best-fit position of an individual source (Mkn~421) and extract its
spectral energy distribution (SED).  Finally the \lstinline$write_roi$
method is used to write the results of the analysis to an output file.

Many of the {\fermipy} methods return a results dictionary that can be
used for subsequent post-processing or visualization.  Results can
also be serialized to a FITS file by passing
\lstinline$write_fits=True$ or a numpy file by passing
\lstinline$write_npy=True$.  The current ROI model can be saved with the
\lstinline$write_roi$ method which generates a FITS file containing a
table with one row per source in the ROI.  This file can be used in
conjunction with the \lstinline$load_roi$ method to restore the
analysis object to a previous state.

%http://fermipy.readthedocs.io/en/\version/output.html

\smallskip
\begin{figure}[t]
\begin{lstlisting}
# Initialize the analysis object
from fermipy.gtanalysis import GTAnalysis
gta = GTAnalysis('config.yaml')
# Setup the analysis
gta.setup()
# Optimize spectral parameters
gta.optimize()
# Localize a source of interest
loc = gta.localize('Mkn421', make_plots=True)
# Extract an SED for a source of interest
sed = gta.sed('Mkn421', make_plots=True)
# Generate a TS map of the region
tsmap = gta.tsmap()
# Save the current analysis state to a file
gta.write_roi('fit0')
\end{lstlisting}
  \caption{An analysis script demonstrating the sequence of methods
    that would be used to perform a basic analysis of an individual
    source (Mkn 421).\label{fig:fermipy_script}}
\end{figure}

\section{Analysis Methods}

{\fermipy} provides a number of high-level analysis methods that
automate common analysis tasks.  The following sections provide 
brief descriptions of some of these methods.

\subsection{\href{http://fermipy.readthedocs.org/en/\version/advanced/sed.html}{sed}}

The {\sed} method computes the SED of a source in the ROI model.  The
only required argument is the name of a source component.  Optional
arguments can be provided to control energy binning and the treatment
of nuisance parameters associated with nearby background sources.  The
SED is computed by replacing the spectral model of the source with a
power law and performing an independent fit of its normalization in
each analysis energy bin.  By default the method will fit the source
normalization using a power law with $\Gamma=2$.  The power-law index
used to fit the source flux in each bin can be controlled with the
\lstinline$bin_index$ and \lstinline$use_local_index$ parameters.

\begin{figure}[t]
\centering
\includegraphics[width=0.49\columnwidth]{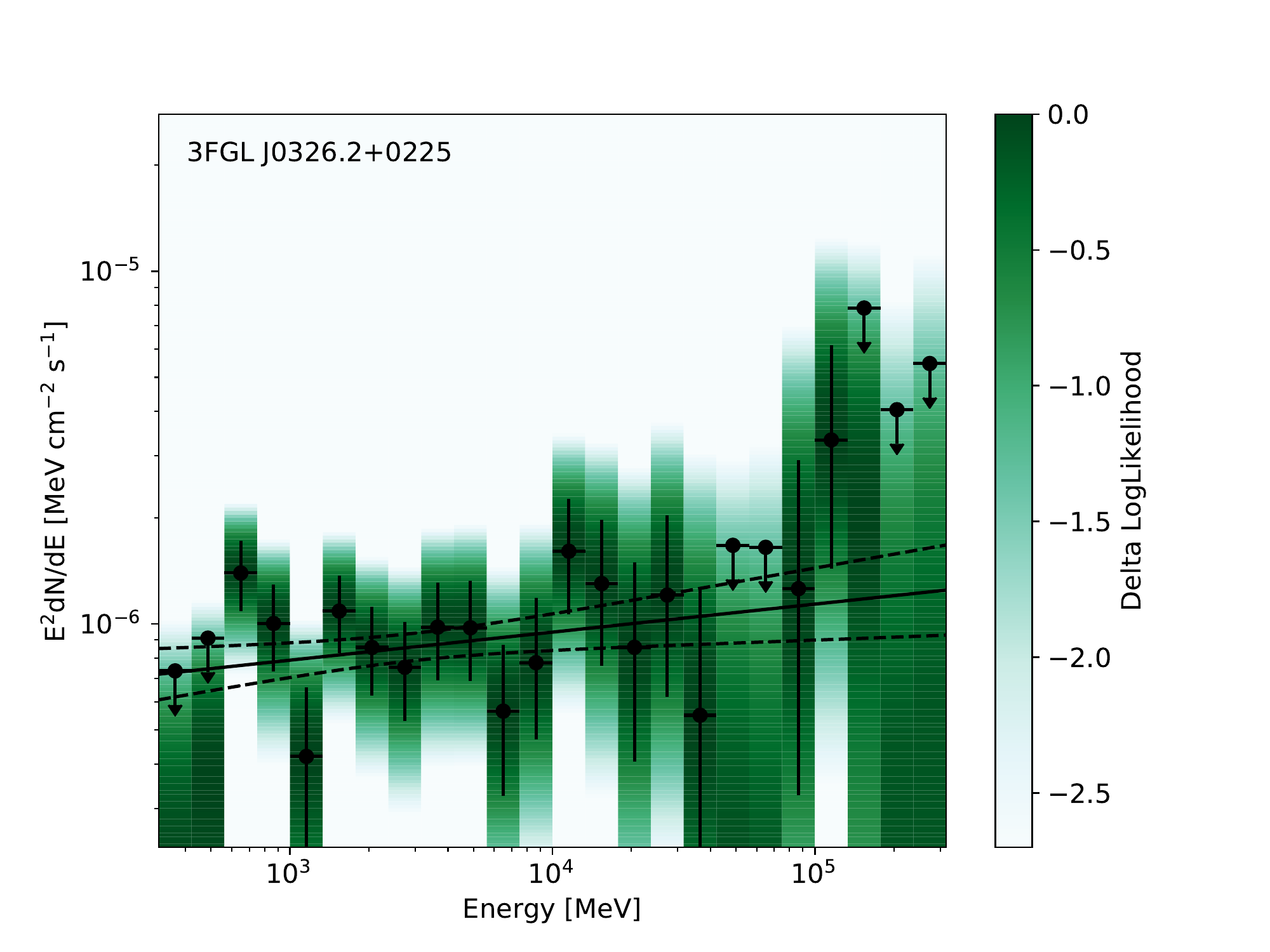}
\includegraphics[width=0.49\columnwidth]{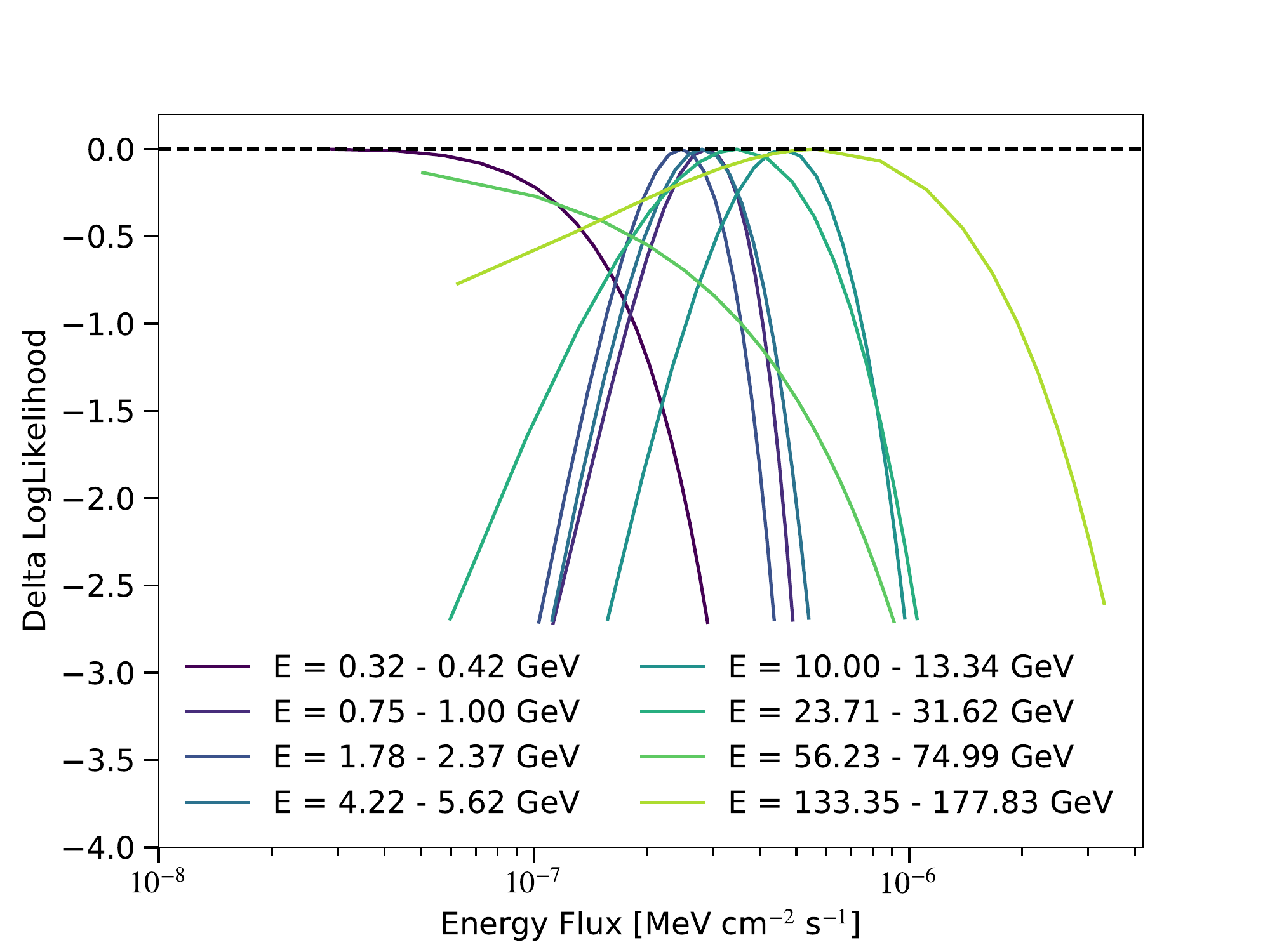}
\caption{\textit{Left:} Diagnostic plot generated by the {\sed} method
  showing the comparison of the spectral points with the best-fit
  global parameterization.  Black points show the best-fit flux or
  95\% flux upper limit measured in each analysis energy bin.  The
  color scale indicates the likelihood profile versus source flux
  extracted in each energy bin. The black solid and dashed lines show
  the uncertainty band (butterfly) of the global
  parameterization. \textit{Right:} Likelihood profiles versus energy
  flux in eight energy bins extracted from the FITS file for the SED
  shown in the left panel.\label{fig:sed} }
\end{figure}

The return value of the method is a dictionary containing the
characteristics of the source in each energy bin: flux, flux errors,
flux upper limits, TS, and predicted counts.  When run with
\lstinline$make_plots=True$, the method generates diagnostic plots
showing the comparison between the measured flux in each bin and the
best-fit global parameterization of the source (see left panel of
Figure~\ref{fig:sed}).  In each energy bin, a profile likelihood
versus source flux is also extracted that can be used to compute flux
confidence intervals under different assumptions or re-fit the global
spectrum using a different spectral parameterization.  The output FITS
file is generated using the SED FITS file format documented at the
\texttt{gamma-astro-data-formats}
repository.\footnote{\url{https://gamma-astro-data-formats.readthedocs.io}}

\subsection{\href{http://fermipy.readthedocs.org/en/\version/advanced/extension.html}{extension}}

\begin{figure}[t]
\centering
\includegraphics[width=0.49\columnwidth]{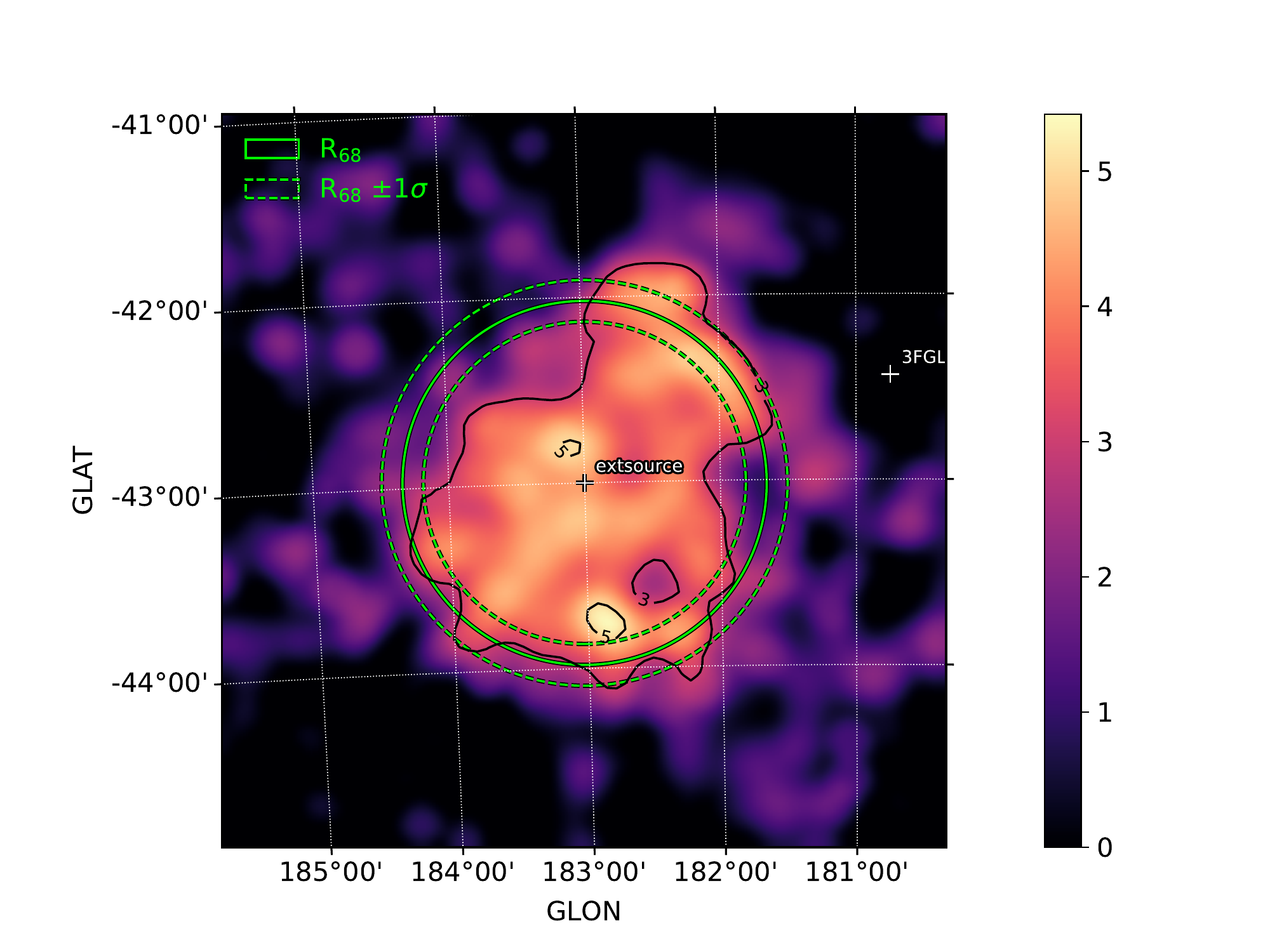}
\includegraphics[width=0.49\columnwidth]{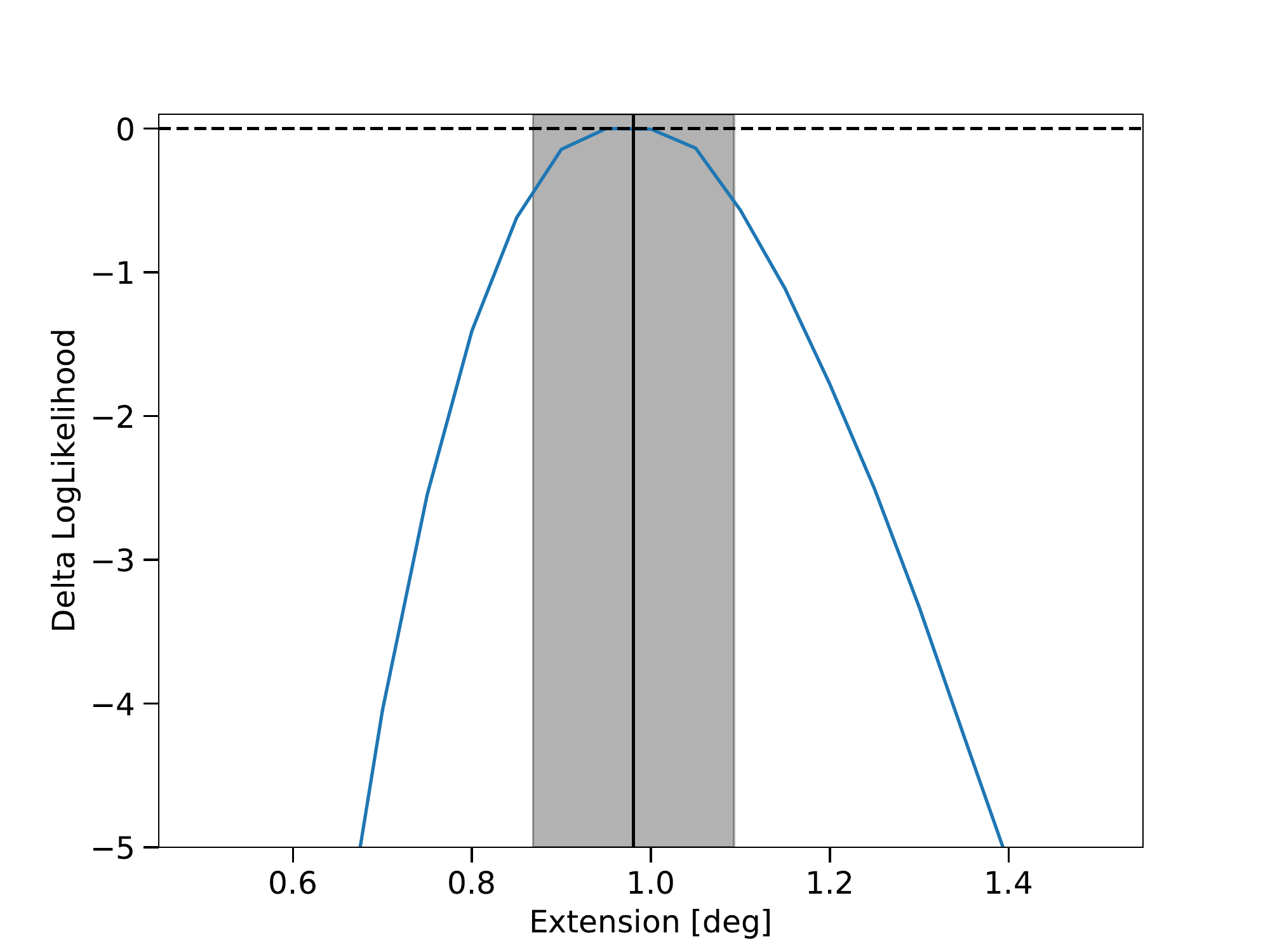}
\caption{\textit{Left:} Diagnostic plot generated by the {\extension}
  method showing the residual TS map of the source overplotted with
  the best-fit extension (green solid and dashed lines).
  \textit{Right:} Likelihood profile versus angular size generated by
  the {\extension} method.  The vertical black line and grey shaded
  region indicate the best-fit value and 1$\sigma$ errors on the
  extension.\label{fig:extension} }
\end{figure}

The {\extension} method can be used to test a source for spatial
extension.  It accepts as its first argument the name of a source and
optional arguments that control the spatial model and whether nearby
background sources are profiled in the fit.  The spatial extension
test is performed by substituting the existing spatial model with an
extended one (e.g. a 2D Gaussian or 2D Disk) and profiling the
likelihood as a function of the spatial size parameter.  The resulting
likelihood profile is used to evaluate likelihood ratio with respect
to a point-source model ($\mathrm{TS}_\mathrm{ext}$), the best fit
extension, and errors and upper limits on the extension.
Figure~\ref{fig:extension} shows some of the diagnostic plots
generated by this method.

\subsection{\href{http://fermipy.readthedocs.org/en/\version/advanced/localization.html}{localize}}

% \begin{figure}[t]
% \centering
% \includegraphics[width=0.49\columnwidth]{{extsource_extension}.pdf}
% \includegraphics[width=0.49\columnwidth]{{extsource_profile}.pdf}
% \caption{\textit{Left:} Diagnostic plot generated by the {\extension}
%   method showing the residual TS map of the source overplotted with
%   the best-fit extension (green solid and dashed lines).
%   \textit{Right:} Likelihood profile versus angular size generated by
%   the {\extension} method.  The vertical black line and grey shaded
%   region indicate the best-fit value and 1$\sigma$ errors on the
%   extension.\label{fig:extension} }
% \end{figure}

The {\localize} method fits the position of a source by maximizing the
model likelihood with respect to source position.  The method first
computes a coarse map of the likelihood as a function of source
position sampled with the analysis pixelization.  A second scan is
then performed using a finer spatial sampling to refine the position
near the peak of the likelihood surface.  The best-fit position and
positional uncertainties are extracted by fitting a 2D
paraboloid to the sampled likelihood values near the peak.  When the
positional fit is complete, the model of the source is updated with
its best-fit position.

%Figure~\ref{fig:extension} shows some of the diagnostic plots
%generated by this method.

\subsection{\href{http://fermipy.readthedocs.org/en/\version/advanced/tsmap.html}{tsmap}}

The {\tsmap} method can be used to rapidly generate a test statistic
map of an ROI using the same algorithm as the \emph{gttsmap} tool.  A
test source is placed at the center of each spatial pixel and a
maximum-likelihood fit is performed to determine its best-fit
amplitude and TS.  The spectral and spatial characteristics of the
test source are controlled with the \lstinline$model$ parameter (by
default a point source with power-law index of 2 will be used).  

\begin{figure}[t]
\centering
\includegraphics[width=0.49\columnwidth]{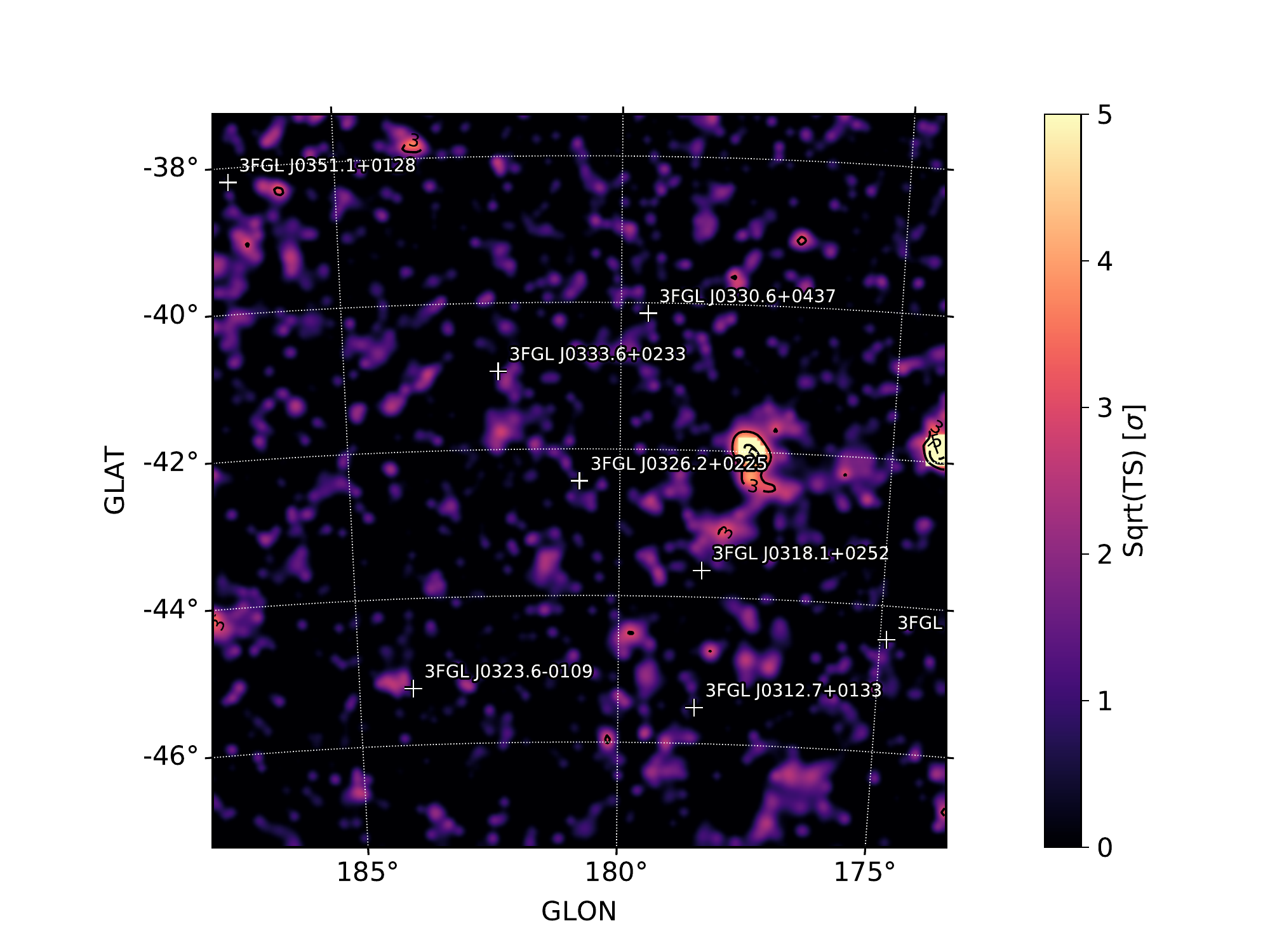}
\includegraphics[width=0.49\columnwidth]{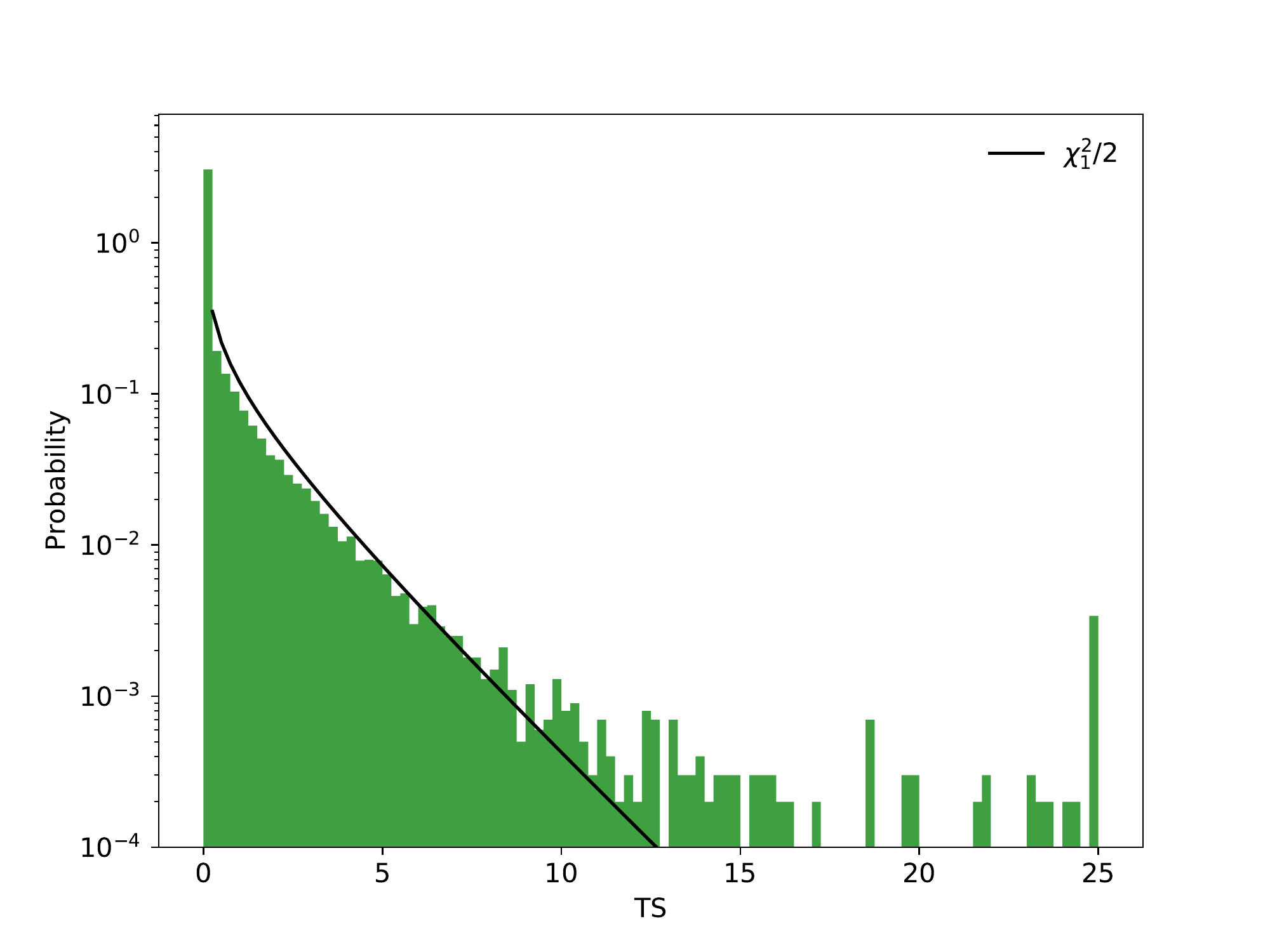}
\caption{Diagnostic plots generated by the {\tsmap} method.
  \textit{Left:} Map of the test-source test statistic versus position
  in the ROI.  White crosses indicate the positions of point sources
  in the model.  \textit{Right:} Histogram of TS values from the TS
  map.  The black curve shows the expected distribution for a
  likelihood ratio test with one degree of
  freedom.\label{fig:tsmap}. }
\end{figure}

To reduce computation time, {\tsmap} fixes the background model and
restricts the calculation to pixels in the vicinity of the test source
position.  These simplifications allow TS maps to be generated much
more rapidly than with the \emph{gttsmap} application.  A TS map of a
typical ROI (e.g. a data cube with $30\times100\times100$ pixels) can
be calculated on a single CPU core in a few minutes.  When run with
\lstinline$make_plots=True$ the method generates a set of diagnostic
plots including those shown in Figure~\ref{fig:tsmap}.  The method
also generates a FITS file containing maps of TS and test-source
normalization.

\subsection{\href{http://fermipy.readthedocs.org/en/\version/advanced/detection.html}{find\_sources}}

The {\findsources} method is an iterative source-finding algorithm
that identifies new source candidates using the likelihood ratio
method.  In each iteration a TS map of the ROI is generated using the
test source model specified with the \lstinline$model$ argument.
Candidate sources are identified from the N highest peaks in the TS
map (where N is set with the \lstinline$sources_per_iter$ parameter)
that have a TS greater than a certain threshold
(\lstinline$sqrt_ts_threshold$) and are at least a certain distance
(\lstinline$min_separation$) from another peak with higher TS.  A new
source is added at each peak satisfying these criteria with a position
derived by fitting a parabola to the TS surface near the peak.  After
each iteration a new TS map is generated using an updated model that
includes the sources added in the last iteration.  The analysis is
stopped when no additional source candidates are found or the number
of iterations exceeds the maximum set with the \lstinline$max_iter$ argument.

\begin{figure}[t]
\centering
\includegraphics[width=0.49\columnwidth]{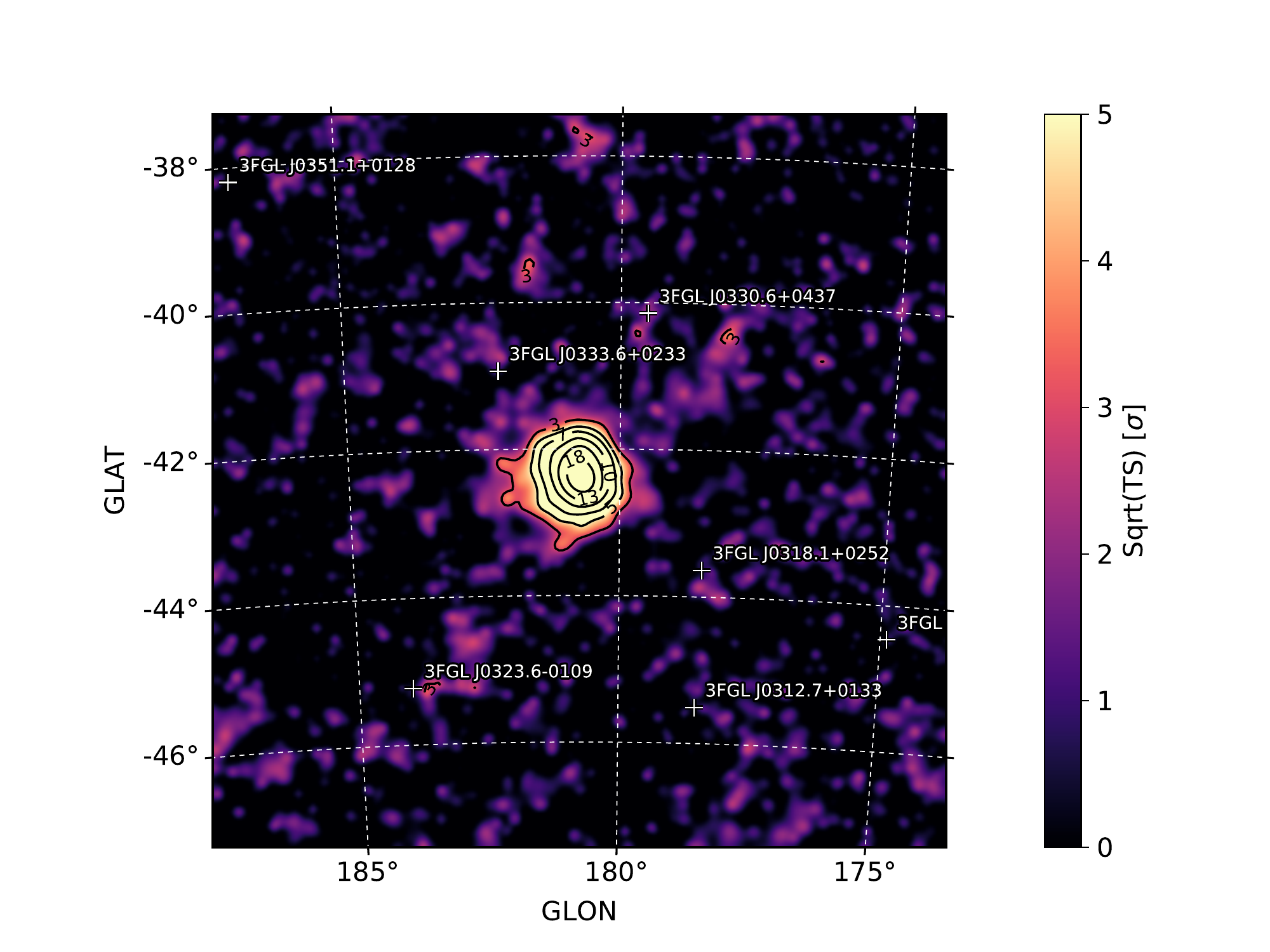}
\includegraphics[width=0.49\columnwidth]{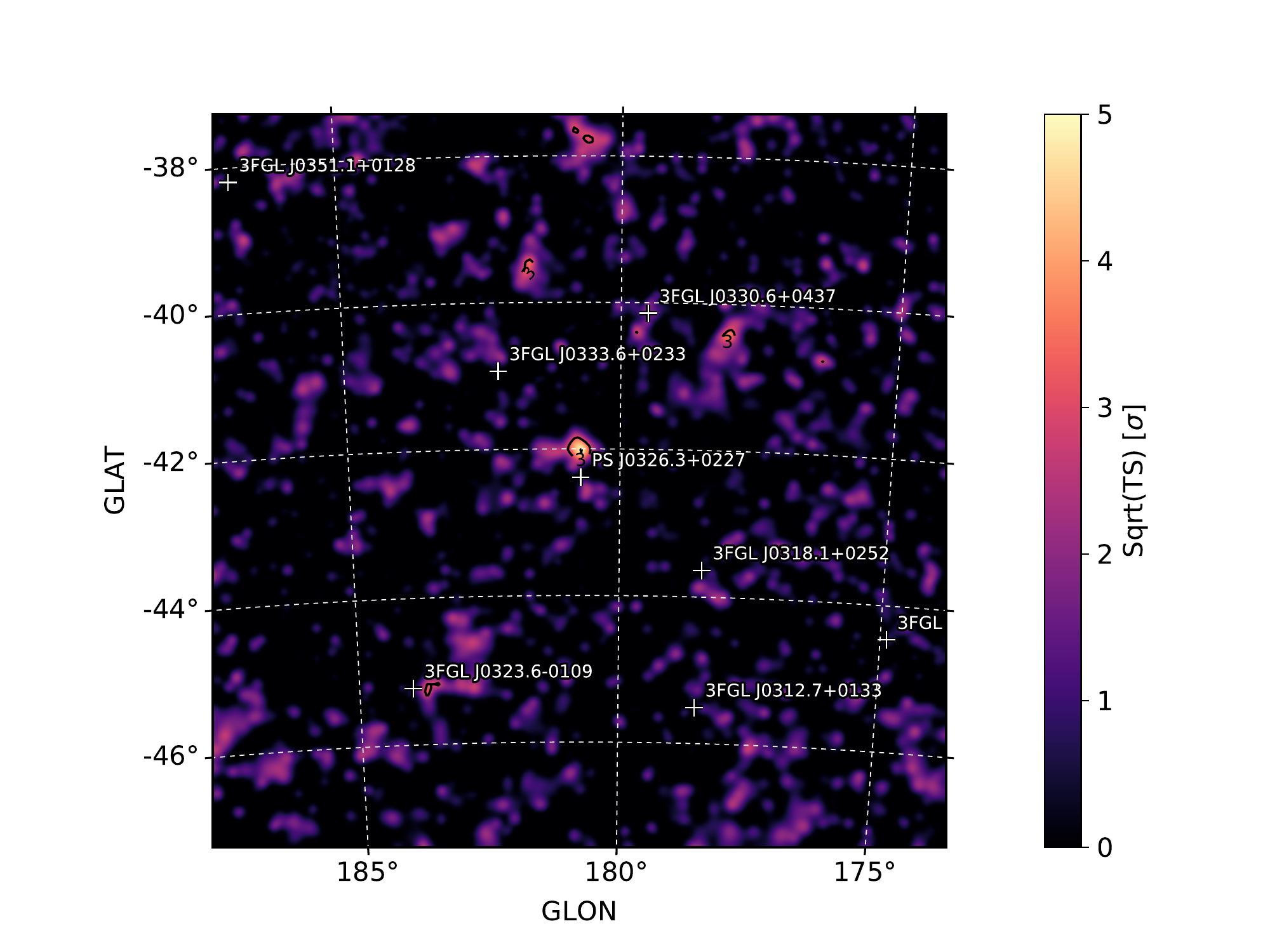}
\includegraphics[width=0.49\columnwidth]{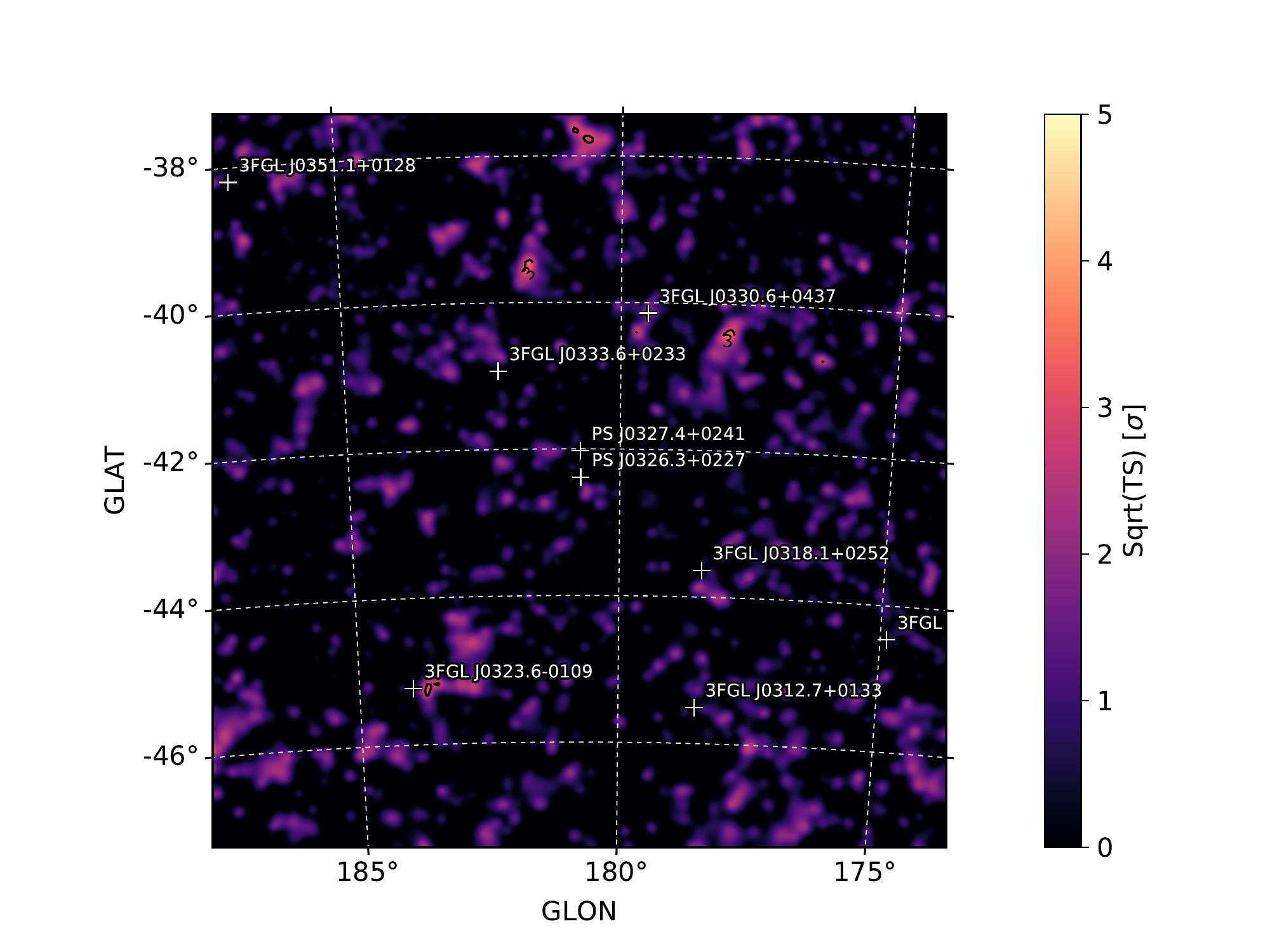}
\caption{Illustration of the {\findsources} method as applied to a
  simulated data set with two unmodeled sources separated by
  $\sim0.5^\circ$.  The top left panel shows the TS map of the ROI
  prior to running the method.  The right and bottom panels show the
  TS maps obtained after the first and second
  iterations.\label{fig:find_sources}. }
\end{figure}

Figure~\ref{fig:find_sources} illustrates the application of this
method to a blended pair of sources.  The first iteration finds a
source in the center of the map.  When this source is added to the
model, a second fainter candidate becomes visible that is identified
as a point source in the second iteration.

\section{Conclusions}

The {\fermipy} package provides an end-to-end tool for high-level
analysis of LAT data that facilitates the study of known gamma-ray
source classes (Active Galactic Nuclei, Pulsars, Supernova Remnants)
as well as searches for new gamma-ray source populations (e.g. Dark Matter
subhalos).
%the diverse gamma-ray source classes of both known and  
%accessible with the LAT, e.g. Active Galactic Nuclei, Pulsars, Supernova
%Remnants, and Dark Matter subhalos.
% facilitates research on a broad range of
% topics (AGN, Galactic sources, DM searches).  
Although the package is now mature, development of {\fermipy} remains
active.  Future releases will focus on the following areas:
\begin{itemize}
\item Support for analyzing data pixelized with HEALPix.
\item Functionality for propagating systematic uncertainties
  associated with the instrument response functions (effective area
  and PSF).
\item Improved integration with other open-source gamma-ray analysis
  software such as \emph{Gammapy}~\cite{2015ICRC...34..789D,gammapy},
  \emph{3ML}~\cite{2015arXiv150708343V}, and
  \emph{Gammalib/ctools}~\cite{2016A&A...593A...1K}.
\item Improved support for multi-threaded analysis.
\item Better standardization of FITS output formats through the
  \texttt{gamma-astro-data-formats} effort.
\end{itemize}
We aim to make {\fermipy} a community-supported effort and to this end
we welcome contributions from all members of the LAT scientific
community.

\acknowledgments

The \textit{Fermi}-LAT Collaboration acknowledges support for LAT
development, operation and data analysis from NASA and DOE (United
States), CEA/Irfu and IN2P3/CNRS (France), ASI and INFN (Italy), MEXT,
KEK, and JAXA (Japan), and the K.A.~Wallenberg Foundation, the Swedish
Research Council and the National Space Board (Sweden). Science
analysis support in the operations phase from INAF (Italy) and CNES
(France) is also gratefully acknowledged.

\bibliographystyle{JHEP}
\bibliography{proceedings}
%\begin{thebibliography}{99}
%\bibitem{...}
%....
%\end{thebibliography}

\end{document}